\documentclass[%
reprint,
superscriptaddress,
nofootinbib,
amsmath,amssymb,
aps,
prl,
floatfix,
longbibliography
]{revtex4-2}

\usepackage{times}
\usepackage{amstext}
\usepackage{graphicx}
\usepackage{color}
\usepackage{soul}
\usepackage{MnSymbol}
\usepackage{subfiles}
\usepackage{bm}
\usepackage{hyperref}
\usepackage{layouts}
\usepackage{enumitem}
\usepackage{natbib}
\usepackage{comment}
\usepackage{xcolor}
\usepackage{multirow}

\begin{document}
\title{Do Venture Capitalists Beat Random Allocation?}

\author{Max Sina Knicker}
\affiliation{Center for Entrepreneurial and Financial Studies, TUM School of Management,
Technische Universität München, Germany}
\affiliation{Econophysics Lab, Institut Louis Bachelier, 28 Pl.~de la Bourse, Palais Brongniart, 75002 Paris, France}

\author{Jean-Philippe Bouchaud}
\affiliation{Econophysics Lab, Institut Louis Bachelier, 28 Pl.~de la Bourse, Palais Brongniart, 75002 Paris, France}
\affiliation{Capital Fund Management, 23 Rue de l'Université, 75007 Paris, France}

\author{Michael Benzaquen}
\affiliation{Econophysics Lab, Institut Louis Bachelier, 28 Pl.~de la Bourse, Palais Brongniart, 75002 Paris, France}
\affiliation{Capital Fund Management, 23 Rue de l'Université, 75007 Paris, France}

\begin{abstract}
\noindent
Venture capital outcomes are dominated by a small number of extreme successes, making it difficult to distinguish investor skill from favorable realizations in a highly skewed return distribution. We study this question by comparing empirical VC portfolios to a constrained random benchmark that preserves key portfolio characteristics, including timing, geography, sector composition, and portfolio size, while randomizing individual company selection.
Across funding stages, empirical portfolio distributions appear remarkably close to their random benchmarks. We find no evidence that portfolio construction increases the probability of high-multiple outcomes: the right tail remains statistically indistinguishable from random allocation. Deviations in the lower part of the distribution are small and sensitive to the interpretation of zero outcomes, suggesting at most weak evidence of downside improvement.
We further introduce a rank-based benchmark distribution to evaluate outperformance at each position in the cross-section. This analysis shows that even the best-performing portfolios do not exceed the outcomes expected for their rank under random sampling. Our results suggest that VC portfolio outcomes are largely consistent with constrained random allocation, highlighting the difficulty of identifying aggregate skill in heavy-tailed investment environments. A similar conclusion holds for the performance of financial analysts in predicting future earnings.
\end{abstract}
\date{\today}

\maketitle

\section{Introduction}

Venture capital returns are highly skewed, with a small number of investments accounting for a disproportionate share of overall performance. This extreme dispersion raises a fundamental question: are top-performing portfolios the result of superior investment skill, or can they largely be explained by favorable outcomes drawn from a highly uneven return distribution? While strong performance is often attributed to investor ability, similar outcomes may arise naturally when returns are dominated by a small number of extreme successes.

Distinguishing between skill and randomness in this setting is inherently difficult. Standard approaches focus on persistence or average outperformance, but these metrics can be hard to interpret in environments where a few extreme outcomes drive aggregate returns. In particular, even in the absence of skill, the best-performing investors may achieve very high returns purely by chance, making it challenging to assess whether portfolio construction systematically improves outcomes.

The question of whether performance reflects skill or randomness has a long tradition in finance and economics, and, more recently, in the context of games and sports (see, e.g., \cite{jerdee2024luck}). Early work on market efficiency argues that abnormal returns are difficult to achieve systematically \citep{jensen1968,carhart1997,fama1970}. A large empirical literature has since focused on decomposing observed performance into skill and luck. A central contribution is \citet{fama2010}, who show that, in mutual funds, most cross-sectional variation in returns can be attributed to noise, with only a small fraction of funds exhibiting statistically significant outperformance. Related work using multiple-testing and bootstrap approaches confirms that true skill is rare and difficult to identify \citep{barras2010,kosowski2006}.

In private equity and venture capital, the evidence is more nuanced. \citet{kaplan2005} and \citet{gompers2006} document persistence in performance, suggesting that skill may exist. However, more recent work highlights the difficulty of separating skill from noise in these settings. In particular, \citet{korteweg2017} show that while variation in fund performance contains a skill component, realized returns are highly noisy, making it challenging to distinguish skilled managers from those who benefit from favorable outcomes. Similar conclusions arise in other domains, where only a small fraction of agents can be reliably identified as skilled, while the majority are statistically indistinguishable from random performance \citep{malladi2020,hartzmark2016}.

A complementary strand of literature emphasizes the role of highly skewed and heavy-tailed return distributions \citep{zajdenweber2000economie, gabaix2009}. When outcomes are dominated by rare extreme events, even large performance differences can arise from randomness rather than systematic ability \citep{bouchaud2000wealth,taleb2007,pluchino2018, bernard2026mean}. This perspective suggests that, even in the presence of skill, aggregate outcome distributions may closely resemble those generated by random processes. Empirical evidence from related settings supports this view. In particular, analyst earnings forecasts exhibit persistent biases and substantial noise \citep{guedj2005experts}, and, as we document in the Appendix, their performance is largely consistent with a random benchmark.

In this paper, we approach this question from a distributional perspective. We compare observed venture capital portfolios to a random benchmark that preserves key structural features, such as timing, sector composition, and geography, while randomizing individual investments. This framework allows us to assess whether portfolio construction systematically shifts the distribution of outcomes, rather than focusing solely on average performance or persistence.

We find that empirical portfolio distributions are close to their random benchmarks across all funding stages, with no evidence of an increased likelihood of high-multiple outcomes. Deviations from the benchmark are limited and concentrated in the lower part of the distribution, but are small in magnitude and sensitive to data limitations.

This approach complements existing work that focuses on average returns or persistence by explicitly characterizing the full distribution of outcomes under a constrained random allocation.

At the investor level, we propose a rank-based benchmark distribution as a measure of true outperformance under random allocation. This framework allows us to evaluate performance relative to the expected outcome at each position in the cross-section, providing a direct comparison between realized outcomes and those implied by chance. Applying this approach, we find that even the best-performing portfolios do not exceed their rank-implied benchmark, highlighting the difficulty of generating outperformance in environments dominated by extreme outcomes.

\section{Data}\label{sec:data}

Venture capital investing is inherently competitive, with success depending on the ability to identify and invest in startups that can generate exceptional returns. To analyze this landscape, we use the Crunchbase dataset~\cite{crunchbase2024}, which provides detailed information on publicly announced funding rounds for startups and their investors. This dataset allows us to characterize the realized investment outcomes of VCs across different funding stages.
The dataset includes detailed records of funding rounds in which a VC participated, the total funding amount for each round, and subsequent funding events for the same startups.

\begin{table}[b!]
\caption{\textbf{Table of Venture Capital data:} The dataset is divided into funding rounds (Series A, B, and C) across six sectors. $N$ denotes the number of startup companies, and $P$ represents the number of investors.}
\begin{tabular}{p{0.38\linewidth}|p{0.05\linewidth}|p{0.13\linewidth}|p{0.13\linewidth}|p{0.13\linewidth}}
                                &  & Series A & Series B & Series C \\\hline
\multirow{2}{*}{Tech \& Software} & $N$ & 2222 & 1108 & 614   \\
                                & $P$ & 1469 & 821 & 500   \\\hline
\multirow{2}{*}{Financial \& Legal} & $N$ & 1187 & 520 & 288   \\
                                & $P$ & 824 & 415 & 248   \\\hline
\multirow{2}{*}{Business \& Services} & $N$ & 3679 & 1751 & 936   \\
                                    & $P$ & 2001 & 1128 & 668   \\\hline
\multirow{2}{*}{Consumer \& Commerce} & $N$ & 3489 & 1515 & 778   \\
                                & $P$ & 1868 & 1040 & 612   \\\hline
\multirow{2}{*}{Health \& Environment} &$N$ & 3128 & 2301 & 1433   \\
                                    & $P$ & 1788 & 1379 & 999   \\\hline
\multirow{2}{*}{Communication \& Ed.} & $N$  & 931 & 398 & 244   \\
                                    & $P$ & 674 & 317 & 218   \\\hline\hline
\multirow{2}{*}{$\sum$} & $N$  & 14636 & 7593 & 4293   \\
                        & $P$ & 8624 & 5100 & 3245
\end{tabular}
\label{tab:vcdata}
\end{table}

We focus on startups indexed by $\alpha$, founded between 2010 and 2022, that received investments from VCs with at least two investments during this period. For each VC $i$, we construct a portfolio $P_i$, comprising all startups they have backed. The average portfolio multiple for a VC is then defined as
\begin{equation}
    M_i= \langle M_{\alpha} \rangle_{\alpha \in P_i},
\end{equation}
where $M_{\alpha}$ is the investment multiple of an individual startup $\alpha$ in the portfolio. Similarly, $n_i = \sum_{\alpha \in P_i}1$ is the number of investments per VC $i$.

The multiple $M_{\alpha}$ for a portfolio company is calculated as the ratio of the funding amount raised in the next round to the funding amount raised in the current round. This serves as a proxy for the startup's growth in investment amount between rounds.\footnote{It is important to note that we do not have access to company valuations, only the funding amounts. To interpret the multiple as a true measure of growth, we assume that the proportion of shares sold remains consistent across consecutive funding rounds.} If a startup fails to raise a subsequent round within three years, the multiple is set to $M_{\alpha}=0$. This definition captures the absence of a follow-on financing event, but does not uniquely identify economic failure, as companies may either cease operations or transition to self-sustaining growth without requiring additional external capital. However, within the venture capital setting, firms are typically financed with the objective of continued growth and eventual exit (e.g., IPO or acquisition), rather than early profitability. As a result, the absence of follow-on financing is more likely to be associated with unsuccessful outcomes, particularly in earlier stages, although this interpretation cannot be verified directly in the data. Moreover, since the data are based on publicly disclosed funding rounds, unsuccessful outcomes are less likely to be observed, which may lead to an under-representation of extreme negative realizations. If multiple VCs invest in the same company during the same round, the company $\alpha$ is included in the portfolios $P_i$ of all participating VCs.

We focus on funding rounds from Seed through Series C. Across all rounds and sectors, the dataset encompasses 26,522 startups and 8,249 unique investors. The mean multiple for individual startups is approximately $\langle M_\alpha\rangle \approx 1.78$, while the mean multiple for investors, representing the average performance of their portfolios, is $\langle M_i\rangle\approx 1.47$ (see Table~\ref{tab:vcdata} for a full breakdown).

This empirical distribution of portfolio outcomes forms the basis for the random benchmark introduced in the next section, where portfolios are reconstructed under the same structural constraints but with randomized investment selection.

\begin{figure}[t!]
    \centering
    \includegraphics[width=1\linewidth]{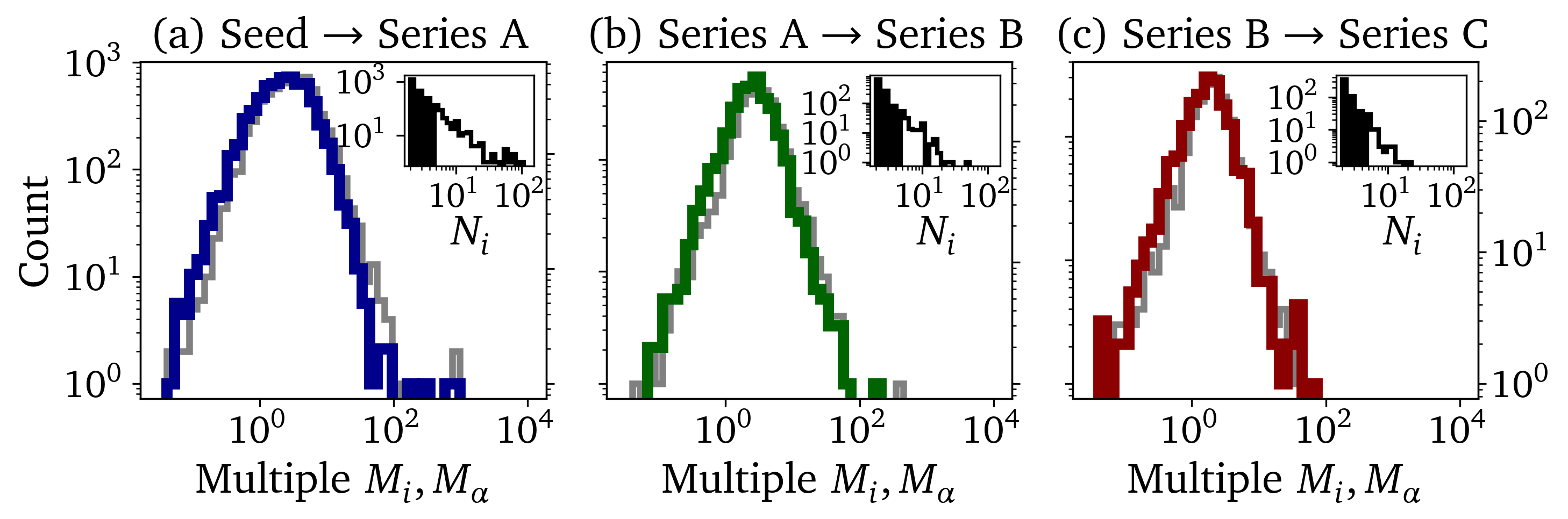}
    \caption{\textbf{Distributions of Venture Capital data.} Distribution of startup multiples (grey) and VC multiples (colored) across funding stages. The distributions are shown for (a) Seed to Series A, (b) Series A to B, and (c) Series B to C. The inset displays the distribution of the number of startup investments per VC. All distributions are shown on a log-log scale.}
    \label{fig:vcdist}
\end{figure}

Figure \ref{fig:vcdist} illustrates the distribution of multiples for both startups (grey) and VCs (colored) across funding stages. The colored distributions represent VC-level outcomes, calculated as the mean multiple of all startups within a VC's portfolio, while the grey distributions represent individual startup multiples. The inset provides the number of startups invested per VC at each stage: Seed $\rightarrow$ Series A (blue), Series A $\rightarrow$ Series B (green), and Series B $\rightarrow$ Series C (red). Both distributions exhibit strong skewness and heavy-tailed behavior, with a small number of startups achieving very high multiples ($M_\alpha \approx 10$), while the majority of outcomes cluster close to or below one. This dispersion highlights that aggregate portfolio outcomes are driven by rare extreme events, a feature that complicates the distinction between skill and randomness.

\section{Benchmarking portfolio distributions}
\label{sec:benchmark}

To assess whether portfolio construction leads to outcomes that differ from random allocation, we compare empirical portfolio multiples against a benchmark generated by \emph{conditional random sampling}. For each investor $i$ with $n_i$ deals in a given stage transition and year, we perform the following Monte Carlo steps:

\begin{enumerate}
  \item For each of the $n_i$ deals, sample (with replacement) a deal from the same calendar year, the same industry cluster, and the same region. This preserves the temporal, sectoral, and geographic composition of $i$'s portfolio but randomizes the individual company exposures.
  \item For each draw and each investor, compute the mean multiple of the sampled deals to obtain a \emph{benchmark multiple} $M_i^{(\mathrm{MC})}$.
  \item Repeat this procedure many times ($1{,}000$ draws) to approximate the distribution of $M_i^{(\mathrm{MC})}$.
\end{enumerate}
Aggregating $M_i$ over all investors yields the \emph{empirical portfolio distribution}, while aggregating $M_i^{(\mathrm{MC})}$ over all investors and all Monte Carlo draws yields the \emph{benchmark distribution}. Since the benchmark respects the size, year, and cluster composition of each portfolio, differences between the empirical and benchmark distributions reflect deviations from what would arise under random allocation within the same opportunity set.

\begin{figure}[t!]
    \centering
    \includegraphics[width=1\linewidth]{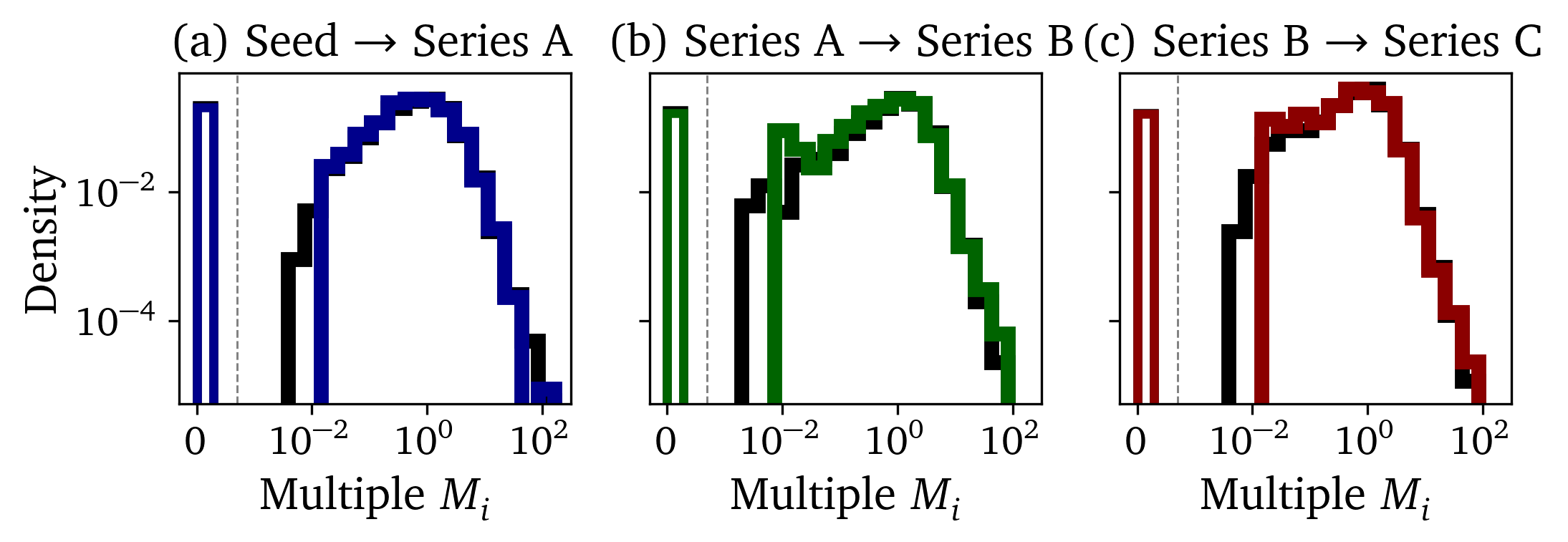}
    \caption{\textbf{Empirical vs.\ benchmark portfolio distributions.} The figure compares the distribution of investor-level portfolio multiples $M_i$ (colored lines) to the benchmark obtained via conditional random sampling (black lines) across funding stages. The lower panels show the log-density difference between empirical and benchmark distributions. See Appendix Fig.~\ref{appx:hist_delta} for linear $y$-scale.}
    \label{fig:hist_delta}
\end{figure}

The plots in Figure~\ref{fig:hist_delta} compare the empirical (colored) and benchmark (black) portfolio distributions on a logarithmic scale. Portfolios with $M_i=0$, corresponding to portfolios in which none of the underlying companies raise a subsequent funding round, are explicitly included in the visualization.\footnote{By construction, the mean of the benchmark distribution matches that of the empirical distribution. On average, approximately 18\% of portfolios result in $M_i=0$, with higher rates observed in early-stage investments and lower rates in later-stage rounds.}

Overall, the empirical distributions are close to their corresponding random benchmarks across all funding stages. While small deviations can be observed in parts of the distribution, these differences are limited in magnitude. In particular, the empirical distribution shows a mild reduction in the frequency of extremely low outcomes ($M \leq 0.1$) relative to the benchmark, accompanied by a slightly lower share of portfolios with $M_i=0$, with a difference of approximately $-2\%$. At the same time, the right tail of the distribution ($M \geq 5$) remains indistinguishable from the benchmark, with no observable increase in the likelihood of high-multiple outcomes.

Taken together, these findings suggest that portfolio outcomes are largely consistent with constrained random allocation: deviations are modest, do not translate into a higher likelihood of high-return outcomes, and much of the observed dispersion can arise from randomness in a highly skewed return environment.

\section{Distribution similarity via Kolmogorov--Smirnov tests}
\label{sec:ks}

\begin{figure}[b!]
    \centering
    \includegraphics[width=1\linewidth]{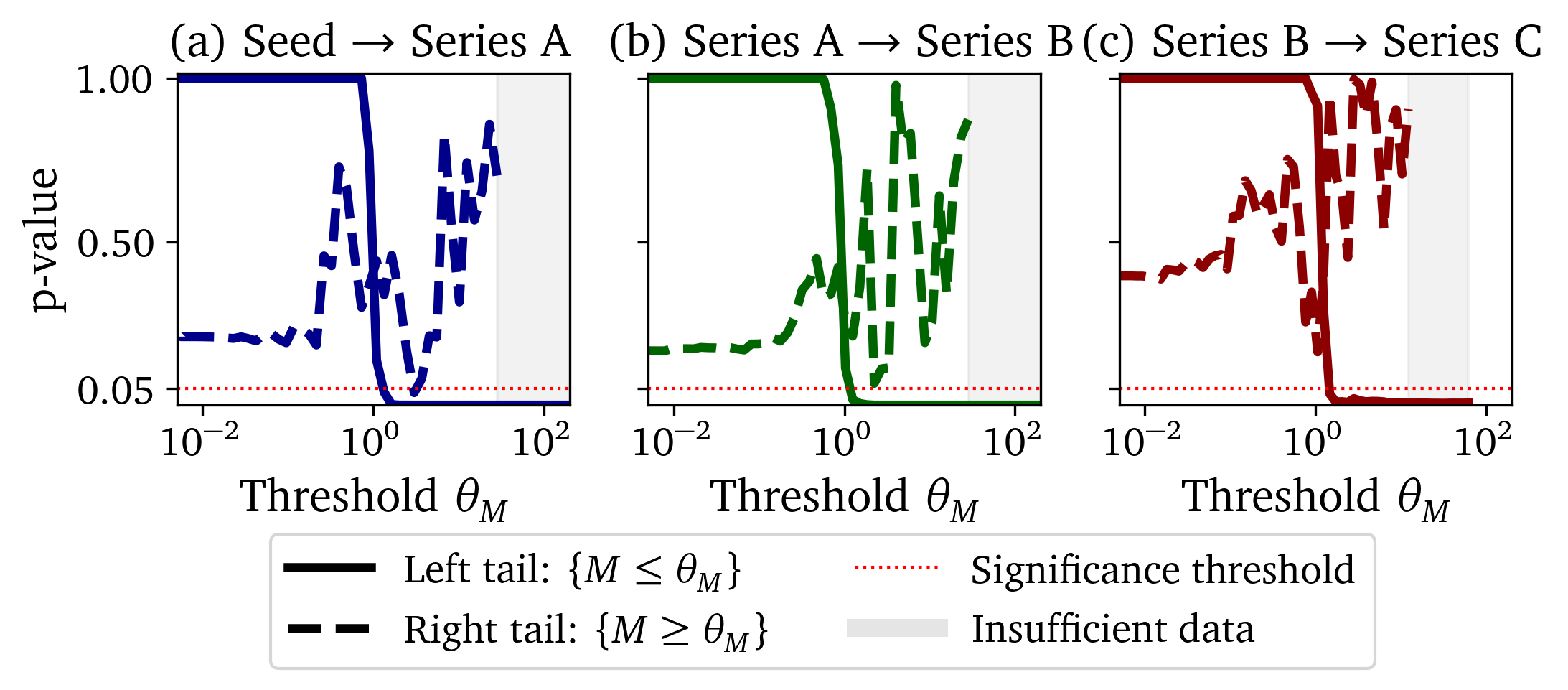}
    \caption{\textbf{Tail-restricted KS tests across the distribution.} The figure shows $p$-values of Kolmogorov–Smirnov tests comparing empirical and benchmark portfolio distributions under left-tail ($M \leq \theta_M$, solid) and right-tail ($M \geq \theta_M$, dashed) restrictions. The red line indicates the $5\%$ significance threshold.}
    \label{fig:ks_sweep}
\end{figure}

To quantitatively assess the similarity between the empirical and benchmark distributions, we conduct Kolmogorov–Smirnov (KS) two-sample tests. The KS statistic measures the maximum vertical distance between the cumulative distribution functions of two samples. A large KS statistic and a small $p$-value indicate that the samples are unlikely to originate from the same underlying distribution.

To distinguish between different regions of the distribution, we consider two one-sided restrictions, rather than computing a single statistic over the full support.
\begin{itemize}
    \item \textbf{Left-tail restricted KS test.} For a set of thresholds $M^\ast$ ranging from the minimum observed multiple to the maximum, we compare the distributions truncated at $M \leq M^\ast$. For very small thresholds, the $p$-values remain high, indicating no statistically detectable difference in the extreme left tail, where observations are sparse (see Figure~\ref{fig:ks_sweep}, solid lines). As $M^\ast$ increases and includes the lower part of the bulk of the distribution, the $p$-value drops below $0.05$, indicating a statistically detectable difference between empirical and benchmark outcomes in this region. This difference remains visible once a sufficiently large portion of the distribution is included.

    \item \textbf{Right-tail restricted KS test.} Similarly, for thresholds $M^\ast$ we compare the distributions truncated at $M \geq M^\ast$. Across most of the range, the $p$-values remain high, indicating that the right tails of the empirical and benchmark distributions are statistically indistinguishable (see Figure~\ref{fig:ks_sweep}, dashed lines). Only at very high multiples do we observe fluctuations, which are not robust and largely driven by sparsity.\footnote{For larger thresholds, the available data become insufficient to reliably compute the KS statistic; these regions are indicated as grey areas in Figure~\ref{fig:ks_sweep}.}
\end{itemize}

The drop of the $p$-value below the $0.05$ threshold in the left-tail restricted KS test occurs once the lower bulk of the distribution is included, indicating that the observed differences are not driven by extreme outcomes alone, but arise within the lower range of typical outcomes. In contrast, the right-tail restricted KS test provides no evidence of statistically significant differences. Overall, these results suggest that any deviations from the benchmark are confined to the lower part of the distribution and are limited in scope, while the upper tail remains statistically indistinguishable from random allocation.

\section{Order statistics and individual outperformance}
\label{sec:orderstats}

While the distributional analysis suggests limited deviations from the benchmark, individual investors could still outperform or underperform relative to a random allocation. To quantify investor-specific deviations, we model the distribution of the $k$-th order statistic of the benchmark portfolio multiples.

Let $\rho(M)$ denote the density of benchmark multiples, and define the cumulative distribution function and survival function as
\begin{align}
    F(M) &= \Pr(M' \leq M), \\
    S(M) &= \Pr(M' > M) = 1 - F(M).
\end{align}

For an investor of rank $k$ (with $k=1$ denoting the highest multiple) among $N$ investors, the probability density that the $k$-th largest multiple equals $M^\ast$ is given, up to normalisation, by the \emph{rank benchmark distribution}
\begin{equation}
  f_{(k)}(M^\ast) \;\propto\; S(M^\ast)^{k-1}\,\rho(M^\ast)\,F(M^\ast)^{N-k},
  \label{eq:rank_density}
\end{equation}
with:
\begin{enumerate}
    \item $S(M^\ast)^{k-1}$: the probability that $k-1$ investors realise multiples above $M^\ast$,
    \item $\rho(M^\ast)$: the density of drawing exactly $M^\ast$,
    \item $F(M^\ast)^{N-k}$: the probability that the remaining $N-k$ investors realise multiples below $M^\ast$.
\end{enumerate}
Eq.~\eqref{eq:rank_density} characterises the benchmark distribution of outcomes associated with a given rank under random selection.

For each stage transition, we estimate $\rho(M)$, $F(M)$, and $S(M)$ from the benchmark sample and construct the corresponding rank benchmark distribution. Normalising this distribution yields
\begin{equation}
  g_{(k)}(M^\ast) = \frac{f_{(k)}(M^\ast)}{\int f_{(k)}(x)\,\mathrm{d}x}.
\end{equation}

From $g_{(k)}(M^\ast)$ we compute the first two moments,
\begin{align}
  \mu_k &= \int M^\ast\, g_{(k)}(M^\ast)\,\mathrm{d}M^\ast, \\
  \sigma_k^2 &= \int (M^\ast - \mu_k)^2\, g_{(k)}(M^\ast)\,\mathrm{d}M^\ast.
\end{align}

We then evaluate the observed performance of the investor at rank $k$ via the standardized deviation
\begin{equation}
  z_k = \frac{M_{(k)}^{\mathrm{obs}} - \mu_k}{\sigma_k},
\end{equation}
where $M_{(k)}^{\mathrm{obs}}$ denotes the observed $k$-th portfolio multiple. A positive $z_k$ indicates that the observed outcome exceeds the benchmark expectation for that rank, while a negative value indicates underperformance.
\begin{figure}[t!]
    \centering
    \includegraphics[width=1\linewidth]{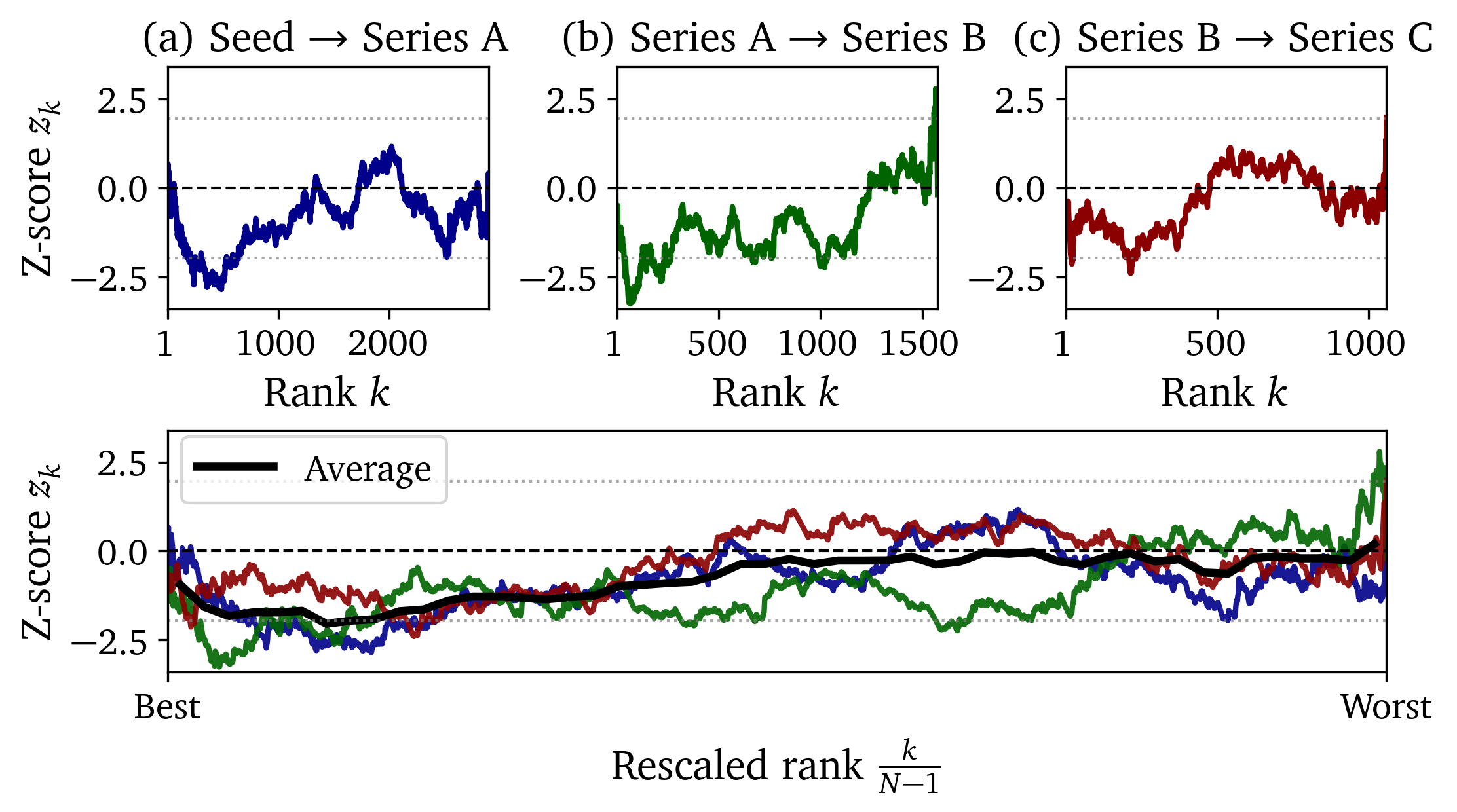}
    \caption{\textbf{Rank-based deviation from the benchmark distribution.} The figure shows the $z$-score $z_k = (M_{(k)}^{\mathrm{obs}} - \mu_k)/\sigma_k$ as a function of investor rank $k$, where $\mu_k$ and $\sigma_k$ are derived from the rank benchmark distribution (Eq.~\eqref{eq:rank_density}). Values above (below) zero indicate outperformance (underperformance) relative to the benchmark expectation for that rank. Top row: disaggregated data; bottom row: same data replotted as a function of $k/(N-1)$ and averaged per bin (thick black line).}
    \label{fig:zscore}
\end{figure}
Figure~\ref{fig:zscore} plots $z_k$ as a function of investor rank (ranked from best to worst) for each stage transition. The lower panel aligns the three rank-based profiles on a common scale by rescaling ranks as ${k}/{(N - 1)}$, mapping performance from best to worst on a scale from 0 to 1. The average is then calculated by binning along the rescaled axis and averaging the corresponding $z$-scores $z_k$ across funding stages.

The rank-based comparison reveals a consistent pattern across stages. At the lower end of the distribution (large $k$), the worst-performing investors cluster around $\langle z_k \rangle \approx 0$, indicating that their outcomes are broadly consistent with what would be expected under random allocation. For the Series A $\rightarrow$ Series B stage, there is some indication that the very worst performers slightly outperform the benchmark, although this effect is not uniform across stages.

At the top of the distribution (small $k$), the best-performing investors tend to exhibit negative values, $\langle z_k \rangle < 0$. This pattern is consistent with the heavy-tailed nature of the benchmark distribution: even under random sampling, the expected top outcomes are already very large, making it difficult for observed portfolios to exceed this benchmark. There is, however, a mild tendency for the very top performers to move closer to the benchmark expectation, suggesting that deviations become less pronounced at the extreme upper end.

Across the entire rank spectrum, the magnitude of $z_k$ remains limited, typically within $\pm 2$, indicating that deviations from the benchmark are small in standardized terms. Overall, these results suggest that portfolio outcomes are broadly consistent with those implied by random allocation, with no evidence of systematic outperformance across the rank distribution. In particular, even the highest-ranked portfolios do not exceed their rank-implied benchmark expectation.

This highlights that extreme performance alone is not sufficient evidence of skill in environments where the distribution of outcomes is highly skewed.

\section{Conclusion}
\label{sec:discussion}

Our analysis compares observed VC portfolio outcomes to a benchmark that preserves the timing, sector composition, geography, and size of each portfolio, while replacing individual investments with random draws. Using Kolmogorov–Smirnov tests, we find that empirical and benchmark distributions are largely similar, with statistically detectable differences confined to the lower part of the distribution. These deviations are limited in magnitude and do not translate into an increased likelihood of high-multiple outcomes. At the distributional level, portfolio outcomes are therefore broadly consistent with those generated by constrained random allocation.

We complement this aggregate view with a rank-based analysis that compares each investor to the outcome expected for that rank under random selection. This perspective leads to a consistent conclusion across the rank spectrum. Investors at the lower end of the distribution perform broadly in line with the benchmark, while even the strongest performers do not exceed the outcomes implied by their rank under random sampling. These results highlight that extreme portfolio performance alone is not sufficient evidence of skill in environments characterized by highly skewed return distributions.

Several caveats are worth emphasising. The analysis is conducted at the aggregated portfolio level and should therefore not be interpreted as definitive evidence that individual managers lack skill or alpha. In particular, our analysis is constrained by data availability. We do not observe the complete set of portfolio companies for all investors, as valuation data are missing for a subset of deals. Multiples are computed based on funding round valuations rather than investor-specific returns, which may differ due to term sheet structures, liquidation preferences, or follow-on participation. Moreover, portfolio-level weights are unobserved, so all investments enter the analysis equally, regardless of their economic size. Finally, outcomes with $M=0$ reflect the absence of a subsequent funding round and may therefore capture both unsuccessful companies and firms that do not return to the funding market, introducing ambiguity in the interpretation of the lower tail.

A second limitation arises from the construction of the benchmark itself. The random sampling procedure assumes that each investor draws from the same year- and cluster-specific distribution as the full sample, thereby abstracting from unobserved heterogeneity such as access, reputation, or network advantages. In addition, the rank benchmark distribution is inherently ex post, as it is constructed from the realised cross-sectional distribution of outcomes. As a result, the benchmark reflects the realised opportunity set and is not directly informative about forward-looking decision-making. Interpreting deviations as evidence of skill therefore requires the implicit assumption that the underlying return distribution is stable over time.

These limitations imply that our results should be interpreted as statements about aggregate outcome distributions, rather than definitive evidence on individual manager skill or investability.

Notwithstanding these limitations, our findings point to a broader conclusion: in environments characterized by heavy-tailed outcomes, it is inherently difficult to generate performance that exceeds what can arise under constrained random allocation. Even the best-performing portfolios do not surpass the benchmark implied by chance, highlighting the challenge of identifying persistent alpha in venture capital markets.

A related analysis of analyst earnings forecasts (see Appendix) reinforces this conclusion, in line with previous work \cite{guedj2005experts}. There, performance is largely indistinguishable from a random benchmark, suggesting that the difficulty of exceeding outcomes implied by chance is not specific to venture capital, but may reflect a more general feature of decision-making under uncertainty.

\section{Acknowledgments}
\label{sec:discussion}

We would like to thank Julius Bonart, Matthieu Cristelli and Stanislao Gualdi for fruitful discussions. 
This research was conducted within the Econophysics
\& Complex Systems Research Chair, under the aegis of
the Fondation du Risque, the Ecole polytechnique and Capital Fund
Management.

\bibliography{refs}

\clearpage

\appendix
\section{Appendix}

\section{Rank Benchmark Distribution}
\vspace{-0.6cm}
\begin{figure}[th!]
    \centering
    \includegraphics[width=1\linewidth]{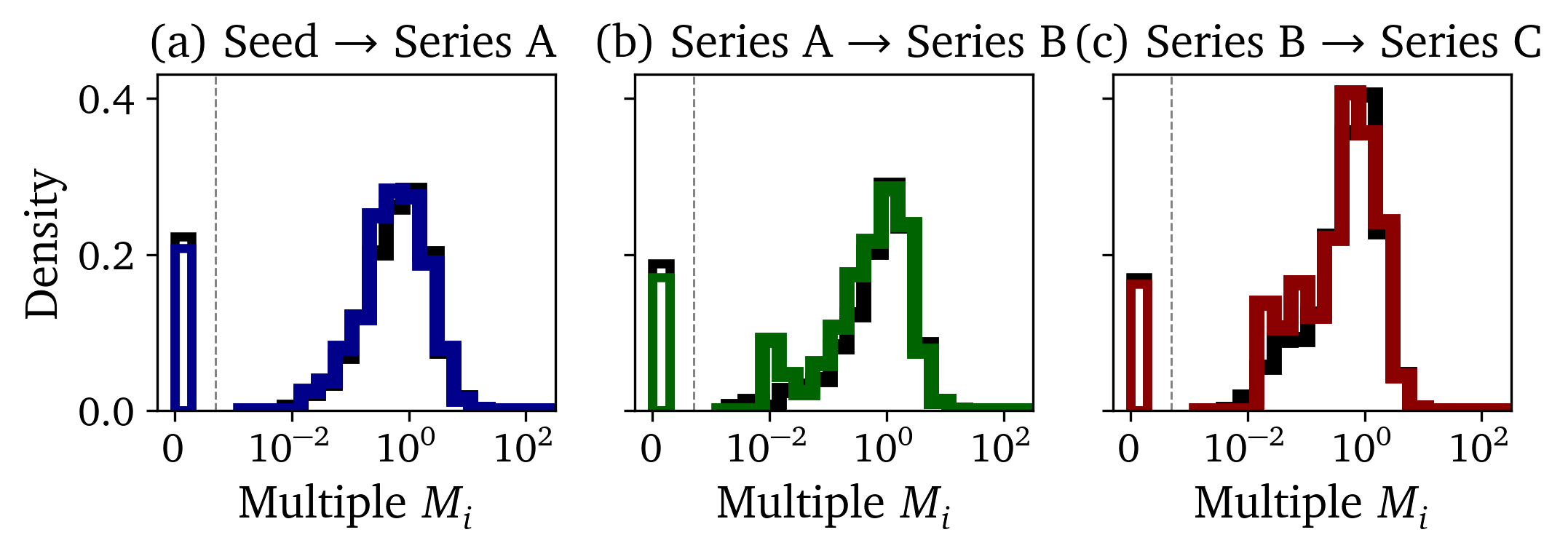}
    \caption{\textbf{Empirical vs.\ benchmark portfolio distributions.} The figure compares the distribution of investor-level portfolio multiples $M_i$ (colored lines) to the benchmark obtained via conditional random sampling (black lines) across funding stages. The lower panels show the log-density difference between empirical and benchmark distributions.}
    \label{appx:hist_delta}
\end{figure}

\begin{figure}[htb!]
    \centering
    \includegraphics[width=1\linewidth]{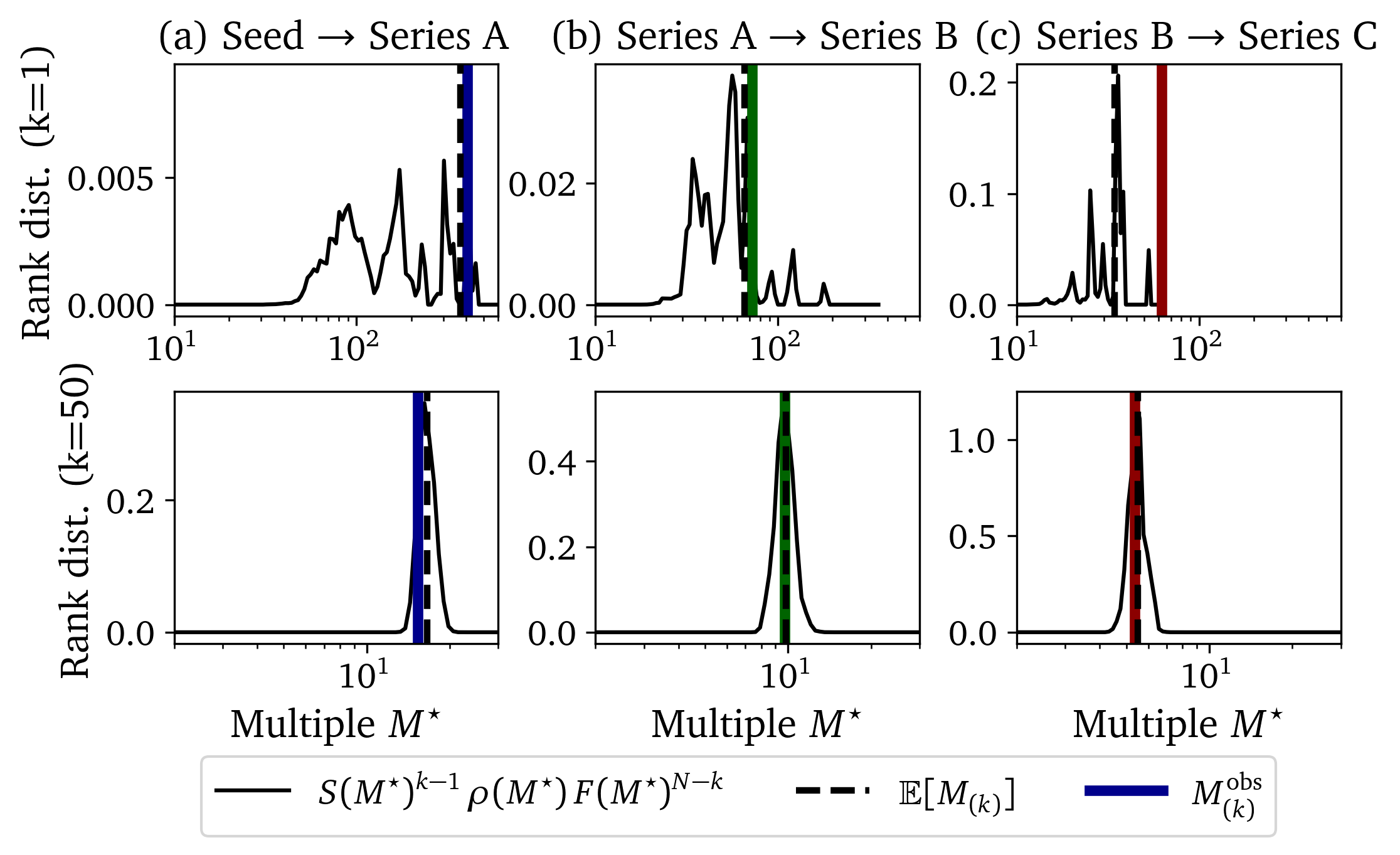}
    \caption{\textbf{Rank benchmark distribution for the top-ranked investor ($k=1$ top, $k=50$ bottom).} The black curve shows the rank benchmark distribution $g_{(1)}(M^\ast)$ implied by Eq.~\eqref{eq:rank_density}, the dashed line its expectation $\mathbb{E}[M_{(1)}]$, and the colored line the observed multiple $M^{\mathrm{obs}}_{(1)}$.}
    \label{fig:appendix_rank}
\end{figure}

Figure~\ref{fig:appendix_rank} illustrates the rank benchmark distribution defined in Eq.~\eqref{eq:rank_density} for two representative ranks, $k=1$ (top performer) and $k=50$. The black curve shows the rank benchmark distribution $g_{(k)}(M^\ast)$ implied by random sampling, while the dashed line indicates its expectation $\mathbb{E}[M_{(k)}]$, and the colored vertical line marks the observed multiple $M^{\mathrm{obs}}_{(k)}$.

For intermediate ranks (e.g.\ $k=50$), the benchmark distribution is relatively concentrated around its mean. In contrast, for the top rank ($k=1$), the benchmark distribution is strongly right-skewed with a high expected value, reflecting the extreme outcomes that can arise from order statistics under a heavy-tailed distribution.

The deviation between $M^{\mathrm{obs}}_{(k)}$ and $\mathbb{E}[M_{(k)}]$ directly corresponds to the standardized score $z_k$ used in the main text.

This representation provides an intuitive interpretation of the rank-based benchmark, illustrating how expected outcomes vary systematically across the cross-section even in the absence of skill.

\section{Analyst Forecasts: A Benchmark Case}\label{sec:forecast}

To put into perspective the findings from venture capital, we consider a related setting in which agents form expectations about firm outcomes but do not directly influence them: equity analyst forecasts. Using the same data as in \cite{guedj2005experts}, we analyse forecasts of annual earnings per share (EPS) across horizons $\theta \in \{0,\dots,11\}$ months prior to the announcement.

To ensure comparability across firms, EPS is normalised by the stock price, yielding an earnings yield $\epsilon_{\alpha,t}$. For analyst $i$, the forecast for firm $\alpha$ at horizon $\theta$ is denoted $f^i_{\alpha,t-\theta}$. We restrict the sample to U.S.\ firms (2012--2019) with at least 10 analysts, resulting in $\approx 4.6\times 10^5$ observations across 2810 analysts.

\begin{figure}[b!]
    \centering
    \includegraphics[width=1\linewidth]{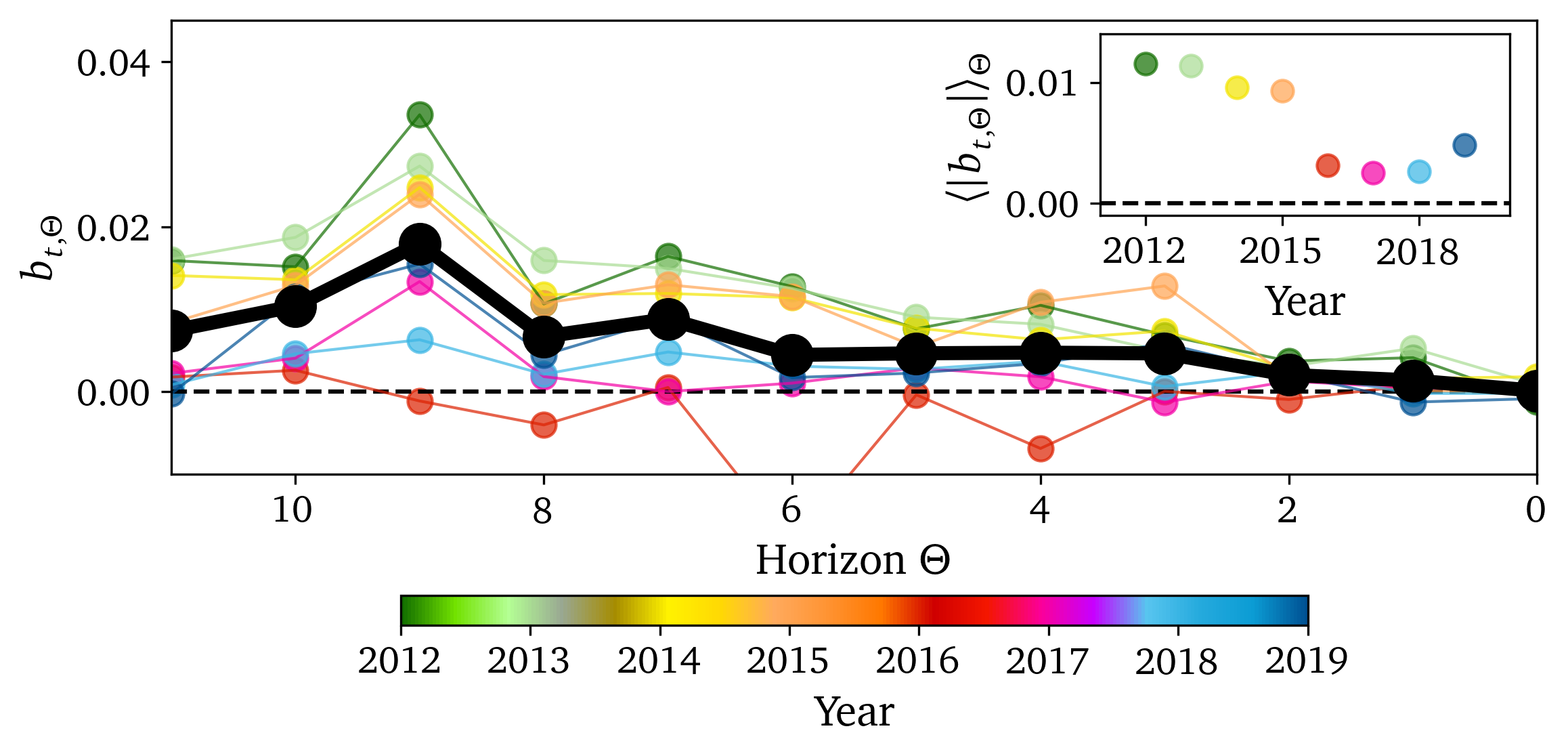}
    \caption{\textbf{Forecast bias across horizons.} The ex-post forecast bias \( b_{\alpha, t-\theta} \) is shown for different years (colored lines) across forecasting horizons \( \theta \), with the black line representing the average bias across all years. Positive values indicate overoptimistic forecasts, also found by \textcite{guedj2005experts} for the period 1987--2004. The inset illustrates average absolute bias across horizons.}
    \label{fig:bias}
\end{figure}
Following \textcite{guedj2005experts}, we define the ex-post forecast bias as
\begin{equation}
    b_{\alpha,t-\theta} = f_{\alpha,t-\theta} - \epsilon_{\alpha,t},
\end{equation}
where $f_{\alpha,t-\theta}=\langle f^i_{\alpha,t-\theta}\rangle_i$ is the consensus forecast.
Figure~\ref{fig:bias} reproduces the key finding of \textcite{guedj2005experts} on more recent data: analyst forecasts exhibit a persistent positive bias, indicating systematic overoptimism. While the magnitude of the bias decreases as $\theta \to 0$, reflecting information arrival, the qualitative pattern remains unchanged, suggesting that forecast optimism is a stable feature of analyst behaviour.

\begin{figure}[htb!]
    \centering
    \includegraphics[width=1\linewidth]{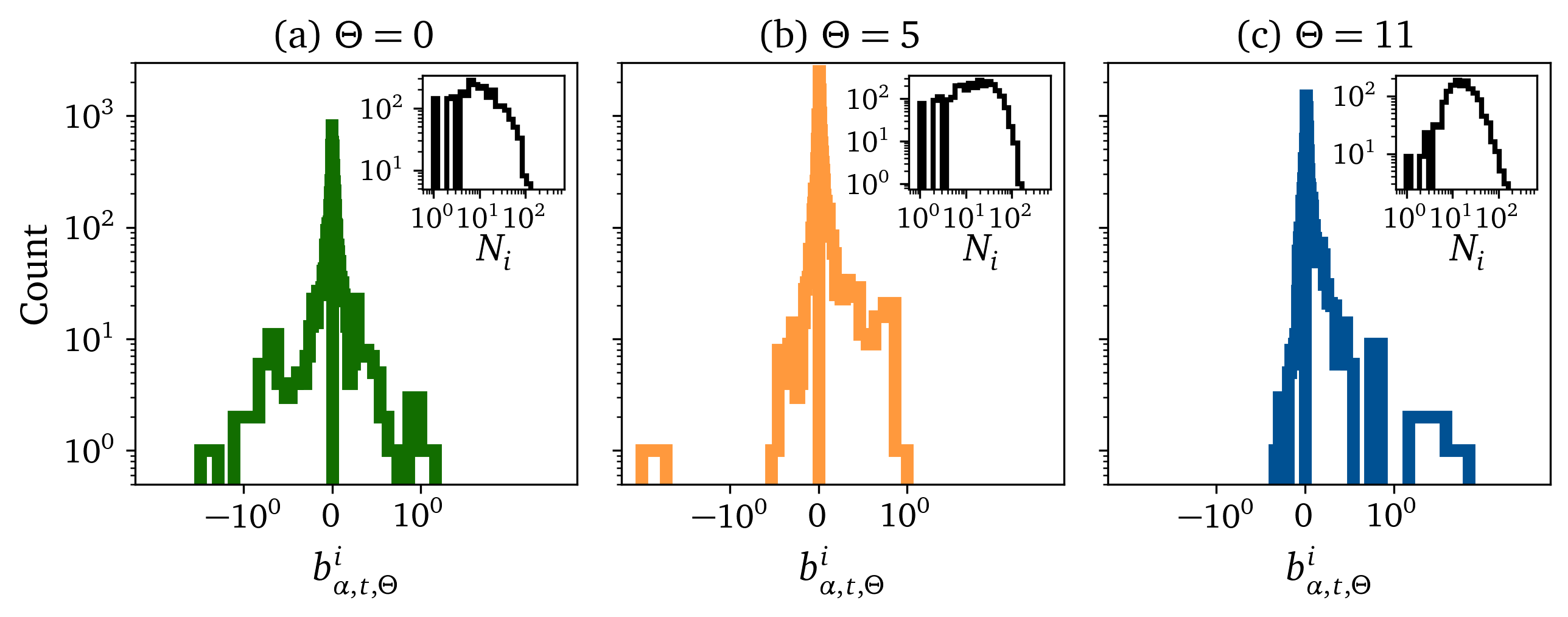}
    \caption{\textbf{Distribution of forecast biases.} The distribution of individual analyst forecast biases \( b^i_{\alpha, t, \theta} \) is shown for horizons \( \theta \in \{0, 5, 11\} \) months before the earnings announcement date. The inset shows the distribution of the number of companies covered per analyst, $N_i$.}
    \label{fig:biasdist}
\end{figure}

Analogous to VC portfolio outcomes, we define the prediction error at the analyst--firm level as
\begin{equation}
    \Delta^i_{\alpha,t-\theta} = \frac{|f^i_{\alpha,t-\theta}-\epsilon_{\alpha,t}|}{\epsilon_{\alpha,t}},
\end{equation}
and aggregate across firms to obtain an analyst-level error,
\begin{equation}
    \Delta^i_{t-\theta} = \langle \Delta^i_{\alpha,t-\theta} \rangle_\alpha.
\end{equation}
Figure~\ref{fig:biasdist} shows that these errors are heavy-tailed, with most predictions close to realised outcomes but a small number of large deviations. The distribution remains skewed at longer horizons and compresses towards zero as $\theta$ decreases.

\begin{figure}[t!]
    \centering
    \includegraphics[width=1\linewidth]{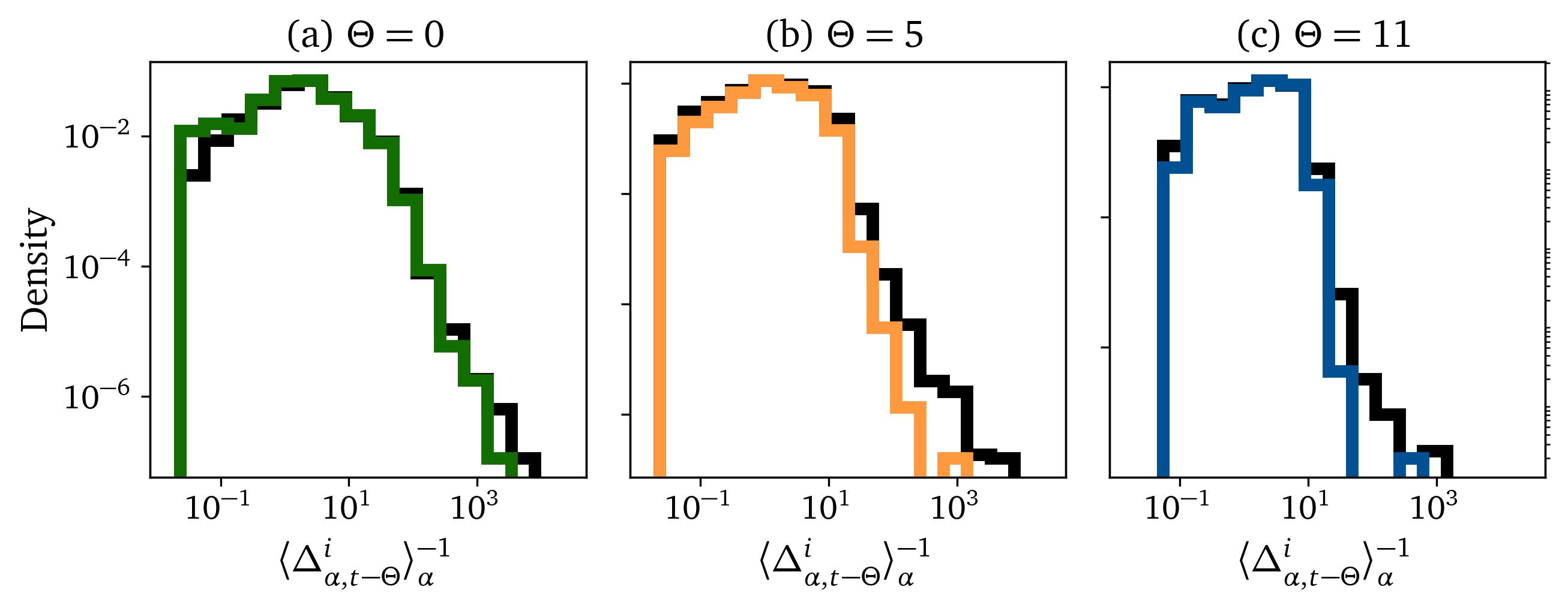}
    \caption{\textbf{Random shuffling of analyst predictions.} The distribution of inverse prediction error $(\Delta^i_{t-\theta})^{-1}$ is shown for actual analyst performance (colored) compared to a random predictor baseline (black) for three forecasting horizons: (a) \( \theta = 0 \), (b) \( \theta = 5 \), and (c) \( \theta = 11 \) months before the earnings announcement.}
    \label{fig:random_shuffling}
\end{figure}
To isolate systematic performance, we construct a random benchmark via conditional reshuffling: for each firm, year, and horizon, analyst errors are randomly reassigned. This preserves the joint distribution of forecasting difficulty while removing persistent analyst-specific effects. Repeating this procedure yields a benchmark distribution of $\Delta^i_{t-\theta}$ under random allocation.

Figure~\ref{fig:random_shuffling} compares the empirical distribution of analyst performance, measured as the inverse prediction error $(\Delta^i_{t-\theta})^{-1}$, to the corresponding random benchmark. With this transformation, higher values indicate better performance, aligning the interpretation with the venture capital setting. The distributions are similar across all horizons. In the right tail, corresponding to the most accurate analysts, observed outcomes fall slightly below the benchmark, particularly at longer horizons, indicating that top performance does not reach the levels implied by random assignment. In the left tail, corresponding to poorly performing analysts, the distributions are largely indistinguishable, with at most a mild tendency toward slightly worse outcomes in the empirical data.

This pattern is consistent with the broader observation that performance in such settings is difficult to distinguish from outcomes generated by random allocation, with only limited and localized deviations.

\begin{figure}[b!]
    \centering
    \includegraphics[width=1\linewidth]{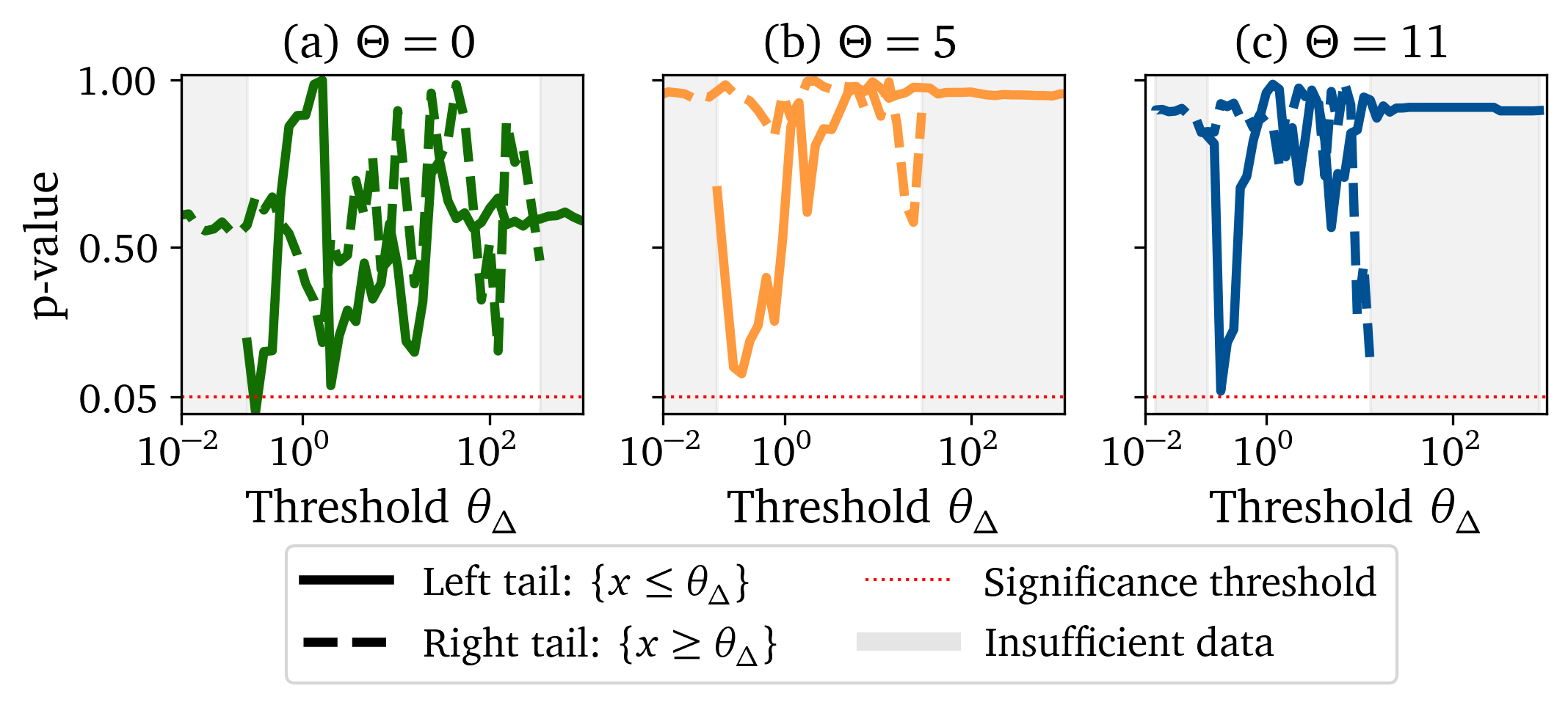}
    \caption{\textbf{Tail-restricted KS tests across the distribution.} The figure shows $p$-values of Kolmogorov–Smirnov tests comparing empirical and benchmark analyst performance distributions under left-tail ($\Delta \leq \theta_\Delta$, solid) and right-tail ($\Delta \geq \theta_\Delta$, dashed) restrictions. The red line indicates the $5\%$ significance threshold.}
    \label{fig:ks_sweep_public}
\end{figure}

The KS results in Figure~\ref{fig:ks_sweep_public} confirm this visual impression. Across thresholds, $p$-values remain high, indicating no statistically significant differences between empirical and benchmark distributions, except for a weak signal in the extreme right tail. Overall, analyst performance appears largely consistent with random allocation across the full distribution.

Taken together, these findings reinforce the main conclusion of the paper: in environments characterized by noise and heavy-tailed outcomes, it is inherently difficult to achieve performance that systematically exceeds what would be expected under random allocation. While venture capital portfolios exhibit small deviations in parts of the distribution, analyst forecasts provide an even clearer example of outcomes that closely track the random benchmark.

\end{document}